\documentclass[twocolumn]{autart}    

\usepackage{amsfonts}
\usepackage{amssymb}
\usepackage{algorithm}
\usepackage{algorithmicx}  
\usepackage{algpseudocode} 
\usepackage{graphicx}          
\usepackage{amsmath}
\usepackage{booktabs} 
\usepackage{epsfig}
 \usepackage{subfigure}
\usepackage{float}
\usepackage{stfloats}
\usepackage{pifont}
\usepackage{color}

\begin{document}

\begin{frontmatter}

\title{Attitude Estimation via Matrix Fisher Distributions on SO(3) Using Non-Unit Vector Measurements \thanksref{footnoteinfo}} 

\thanks[footnoteinfo]{This note was not presented at any IFAC 
meeting. This work was supported in part by the National Natural Science Foundation
of China under Grant 12422214, National Key R\&D Program of China (No. 2024YFB3909900
), University-Research Cooperation Fund of the Eighth Research Institute of China Aerospace Science and Technology Corporation under Grant SAST2024-061 and the Taihu Lake Innovation Fund for Future Technology.}

\author[Shenyuan]{Shijie Wang}\ead{wangsj2001@buaa.edu.cn},    
\author[Astronautics,Key]{Haichao Gui}\ead{hcgui@buaa.edu.cn},               
\author[Astronautics]{Rui Zhong}\ead{zhongruia@163.com}  

\address[Shenyuan]{School of Astronautics, Shenyuan Honors College, Beihang University, Beijing, 100191, China}
\address[Astronautics]{School of Astronautics, Beihang University, Beijing, 100191, China}
\address[Key]{Key Laboratory of Spacecraft Design Optimization \textit{{\rm \&}} Dynamic Simulation Technologies, Ministry of Education, Beijing, China}

\begin{keyword}                           
Matrix Fisher distributions; Bayesian estimation; Attitude estimation; Nonlinear observer and filter design               
\end{keyword}                             

\begin{abstract}                          
This note presents a novel Bayesian attitude estimator with the matrix Fisher distribution on the special orthogonal group, which can smoothly accommodate both unit and non-unit vector measurements. The posterior attitude distribution is proven to be a matrix Fisher distribution with the assumption that non-unit vector measurement errors follow the isotropic Gaussian distributions and unit vector measurements follow the von-Mises Fisher distributions. Next, a global unscented transformation is proposed to approximate the full likelihood distribution with a matrix Fisher distribution for more generic cases of vector measurement errors following the non-isotropic Gaussian distributions. Following these, a Bayesian attitude estimator with the matrix Fisher distribution is constructed. Numerical examples are then presented. The proposed estimator exhibits advantageous performance compared with the previous attitude estimator with matrix Fisher distributions and the classic multiplicative extended Kalman filter in the case of non-unit vector measurements.
\end{abstract}

\end{frontmatter}

\section{Introduction}
Attitude estimation has been studied with the measurement of angular rate gyros and body-frame observations of reference vectors since the 1960s \cite{crassidisSurveyNonlinearAttitude2007b}. One of the major challenges in dealing with the uncertainty in the estimation is that the attitude of a rigid body evolves on a nonlinear, compact manifold, which is referred to as the special orthogonal group $SO(3)$. The multiplicative extended Kalman filter (MEKF) with quaternions \cite{leffertsKalmanFilteringSpacecraft1982} \cite{markley2003attitude} is a milestone in the development of attitude estimation, which avoids the singularities caused by three-parameter attitude representations. More recently, the invariant extended Kalman filter (IEKF) was developed with more geometric properties of attitude system introduced \cite{bonnabelLeftinvariantExtendedKalman2007,barrauInvariantKalmanFiltering2018,phogatInvariantExtendedKalman2020} and was applied to 
the attitude estimation of spacecraft \cite{guiQuaternionInvariantExtended2018}. Despite the success that the MEKF and IEKF methods have achieved, Gaussian distributions on Euclidean space are used to locally represent the uncertainty of attitude and the resulting errors are not negligible especially when attitude uncertainty is large \cite{wang2021matrix}. The importance of defining probability distributions properly on the Lie groups for attitude estimation has been increasingly noticed among the robotics community \cite{barrauInvariantKalmanFiltering2018}.

Several recent attempts have been made to construct estimators with probability distributions directly defined on $SO(3)$. There have been efforts to apply the directional statistics \cite{mardiaDirectionalStatistics1999}, a branch of statistics that studies random variables and matrices on compact manifolds, to attitude estimation. The matrix Fisher distribution, introduced in \cite{mardiaDirectionalStatistics1999}, is a compact form of an exponential density model provided by directional statistics. Based on the matrix Fisher distribution on $SO(3)$, a Bayesian estimator was constructed in \cite{leeBayesianAttitudeEstimation2018} and further developed in \cite{wang2021matrix} and \cite{wangObservabilityAttitudeSingle2022}.  

Although a series of works on Bayesian estimators with the matrix Fisher distribution showed particular advantages in dealing with large initial errors and measurement uncertainty, these filters narrowly constrain the noise model of vector measurements. In \cite{leeBayesianAttitudeEstimation2018}, \cite{wang2021matrix}, and \cite{wangObservabilityAttitudeSingle2022}, the vector measurements are assumed to be unit and to follow the von-Mises Fisher distribution \cite{ley2018applied} on the two-dimensional sphere $S^2$. The assumption ensures that the posterior distribution is a matrix Fisher distribution. However, the unit assumption of vector measurement does not apply to some widely used sensors, such as magnetometers and accelerometers. For non-unit vector measurements, additive errors characterized by Gaussian distributions are used to model sensor noise. Directly normalizing non-unit vector measurements and approximating the isotropic additive Gaussian noise with the von-Mises Fisher distribution for using these estimators can cause loss of accuracy \cite{chave2015note}. Moreover, only one scalar parameter in the von-Mises Fisher distribution is defined to describe the dispersion of normalized vectors. Therefore, it is difficult to characterize non-isotropic stochastic errors. Therefore, it is more direct and precise to design attitude estimators via the matrix Fisher distribution that can accommodate non-unit measurements without normalization, which has not been addressed so far and is considered in this note.

In this note, we design a Bayesian estimator for attitude estimation with the matrix Fisher distribution, which can accommodate both unit and non-unit vector measurements. The proposed estimator maintains the advantages over the classic MEKF. Meanwhile, compared to the previous estimator with the matrix Fisher distribution, it can more accurately deal with non-unit vector measurements without enforcing normalization, while  keeping essentially the same computational cost. Therefore, the proposed filter is more cost-effective when vector measurements are non-unit. In the proposed estimator, vector measurements with isotropic and non-isotropic additive Gaussian noise are treated respectively with different methods. For the former, we prove that the posterior distribution of attitude is a matrix Fisher distribution when non-unit vector measurements follow the isotropic Gaussian distribution and unit vector measurements follow the von-Mises Fisher distribution in Theorem \ref{th:iso_postri}. This theorem extends the application scope of Theorem 3.2 in \cite{leeBayesianAttitudeEstimation2018} to more generic cases. For the latter, a novel unscented transformation is proposed to propagate uncertainty between the 3-dimensional Euclidean space and $SO(3)$, without using any local coordinates. Unlike the related works in \cite{kurz2016unscented} and \cite{lee2016global}, the proposed unscented transformation propagates uncertainty between two different manifolds instead of the same manifolds. Numerical simulations show that the proposed estimator demonstrates an advantageous performance compared with the MEKF and the estimator presented in \cite{leeBayesianAttitudeEstimation2018} in challenging circumstances.

\section{Preliminaries}
\subsection{Notations and Relevant Properties}
Throughout this note, the $n$-dimensional Euclidean space is denoted by $\mathbb{R}^n$, and the $n\times n$ identity matrix is denoted by $I_{n}$. The three-dimensional orthogonal group $O(3)$ is defined as  $O(3) = \left\{ {R \in \mathbb{R}^{3 \times 3} |RR^T  = I_{n} } \right\},$
which consists of two disconnected subsets with different determinants. The component with a determinant of $+1$ is referred to as the 3-dimensional special orthogonal group and is denoted by $SO(3)$. The Lie algebra of $SO(3)$ is denoted by $\mathfrak{so}(3) = \left\{ { X \in \mathbb{R} ^{3\times 3} | X = -X^T } \right\}$,  and the map $\wedge$ is an isometry between $\mathfrak{so}(3)$ and $\mathbb{R} ^{3}$ and $\vee$ denotes its inverse map. The trace of a matrix is denoted by $\text{tr}(\cdot)$ and the function $e^{\text{tr}(\cdot)}$ is abbreviated as $\text{etr}(\cdot)$. The symmetric matrix $M=M^T$ is positive definite if all eigenvalues are positive, denoted by $M \succ 0$.
The first moment of the random variable $x$ is denoted by $E(x)$. A random variable $x \in \mathbb{R}^n$ following the Gaussian distribution with mean $\mu \in \mathbb{R}^n$ and covariance matrix $C \in \mathbb{R}^{n \times n}$ is denoted by $x \sim \mathcal{N}(\mu,C)$.

The following equations will be used repeatedly. For any $a,b\in \mathbb{R}^3$ and $A,B,C \in \mathbb{R}^{3 \times 3}$,
\begin{equation}
\text{tr}(ab^T)=a^Tb\nonumber,
\end{equation}
\begin{equation}
\text{tr}(ABC)=\text{tr}(C^TB^TA^T)=\text{tr}(CAB)=\text{tr}(BCA)\nonumber.
\end{equation}
\subsection{Matrix Fisher Distributions on SO(3)}
The matrix Fisher distribution was originally developed for the Stiefel manifold, and more recently, has been applied to the special orthogonal group to represent global uncertainty in attitude estimation. The probability density function (PDF) of a random rotation matrix $R \sim \mathcal{M}(F)$ is defined as
\begin{equation}
    p(R\in SO(3);F) = \frac{1}
{{c\left( F \right)}}\text{etr}\left[ {F^T R} \right],\label{eq:MF_de}
\end{equation}
where $F \in \mathbb{R}^{3\times3}$ is a parameter and $c(F)\in \mathbb{R}$ is the normalized constant. The normalized constant $c(F)$ is calculated by $c(F) = \int_{SO(3)} {\text{etr}\left[{F^TR}\right]dR = c(F)}$, where $dR$ is a bi-invariant Haar measure. The parameter $F$ can always be decomposed via proper singular value decomposition (SVD) \cite{markleyAttitudeDeterminationUsing1988} as follows
\begin{equation}
    F=USV^T,\label{eq:SVD}
\end{equation}
where $U,V \in SO(3)$ and $S = \text{diag}[s_1, s_2, s_3]\in \mathbb{R}^{3\times3}$ with $s_1 \geq s_2 \geq |s_3| \geq 0$. Given a random rotation matrix $R \sim \mathcal{M}(F)$, the first moment of $R$ is given by \cite{leeBayesianAttitudeEstimation2018}
\begin{equation}\label{eq:ER}
    E(R) = U\left( \frac{1}{{c(S)}} \text{diag}\left[ \frac{{\partial c(S)}}{{\partial s_1 }}, \frac{{\partial c(S)}}{{\partial s_2 }}, \frac{{\partial c(S)}}{{\partial s_3 }}\right] \right)V^T.\nonumber
\end{equation}
The mean attitude of the matrix Fisher distribution is defined as the attitude that maximizes the density function as well as minimizes the mean squared error, and is given by $M = UV^T \in SO(3)$.

\subsection{System Models}
Consider a rigid body with its body-fixed frame denoted by $\mathcal{F}_B$. Then the attitude of the rigid body is the orientation of $\mathcal{F_B}$ relative to $\mathcal{F_I}$ expressed in $\mathcal{F_I}$, which can be represented by a rotation matrix $R\in \mathbb{R}^{3\times 3}$ with its $(i,\ j)$-element given by $R_{ij}=e_i\cdot b_j$, where $e_i$ is the $i$-th standard base of $\mathcal{F_I}$ and $b_j$ is the $j$-th standard base of $\mathcal{F_B}$.  When both $\mathcal{F_B}$ and $\mathcal{F_I}$ are right-handed orthonormal frames, we have $R\in SO(3)$.
The attitude kinematics model, which describes the attitude evolution on $SO(3)$ and propagates uncertainty of attitude, is considered as \cite{leeBayesianAttitudeEstimation2018},
\begin{equation}\label{eq:c_ki}
\left[{R^T(t) \dot R(t)}\right]^{\vee} = \Omega(t) dt + HdW,
\end{equation}
where the vector $\Omega (t) \in \mathbb{R}^3$ represents the angular velocity with respect to ${\mathcal{F}}_I$ expressed in ${\mathcal{F}}_B$, and it is measured by rate gyroscopes. The vector $W \in \mathbb{R}^3$ is a three-dimensional Wiener process and the matrix $H \in \mathbb{R}^{3 \times 3}$ describes the strength of the measurement noise. We assume that time is discretized by a sequence $\{{t_1,t_2,\dots,t_k,\dots}\}$ with a fixed time step $t_{k+1}-t_k \equiv h$, and that $\Omega(t)$ remains constant within each time interval $[t_k,t_{k+1}]$. The equation (\ref{eq:c_ki}) is then discretized into
\begin{equation}\label{eq:d_ki}
R_{k+1} = R_k \exp\left\{{h \Omega_k + (H_k \Delta W_k)^{\wedge}}\right\},
\end{equation}
where $\Omega_k = \Omega(t_k)$ and $\Delta W_k$ is the stochastic increment of the Wiener processes over a time step, which is a Gaussian with $H_k \Delta W_k \sim \mathcal{N}(0_{3 \times 1},hG_k)$, where $G_k = H_k{H_k}^T$.  Meanwhile, the analytic expression of the first moment of attitude following the matrix Fisher distribution along (\ref{eq:d_ki}) is
\begin{align}\label{eq:MF_proga}
			E\left( {R_{k + 1} } \right) &= E\left( {R_k } \right)\left\{ {I_{3 \times 3}  + 0.5h\left[ { - \text{tr}\left( {G_k } \right)I_{3 \times 3} } \right.} \right. \nonumber\\ 
			&\left. {\left. { + G_k } \right]} \right\}\exp (h\Omega _k ) + \mathcal{O}(h^{1.5} ).
	\end{align}
 
where $\mathcal{O}(h^{1.5} )$ includes all terms of order at least $h^{1.5}$. Denote the $i$-th reference vector at $t_k$ as $r_k^i \in \mathbb{R}^3$ in $F_{\mathcal{I}}$ and the corresponding measurement as $z_k^i \in \mathbb{R}^3$ in $F_{\mathcal{B}}$. The measurement model for attitude sensors is given by
\begin{equation}\label{eq:mea_model}
z_k^i = R_k^Tr_k^i+v_k^i,
\end{equation}
where $v_k^i\in \mathbb{R}^3$ represents the random error, which is zero-mean Gaussian with
the covariance matrix $Q_k^i \in \mathbb{R}^{3 \times 3}$, such that $Q_k^i =(Q_k^i)^T \succ 0$. Vector measurements are assumed to be independent of each other, i.e.
\begin{equation}\label{eq:mea_cov}
E\left[v_k^i(v_k^j)^T\right] = 0_{3 \times 3},\ for\ i \ne j.
\end{equation}
 
\section{Bayesian Estimator with the Matrix Fisher Distribution}
\label{sec:4}

\subsection{The Posterior Distribution for the Isotropic Gaussian Distribution}\label{sec:4_1}
To construct the attitude Bayesian estimator with the matrix Fisher distribution, we first show the relation between the posterior attitude probability distribution and the matrix Fisher distribution. The following theorem can accommodate both unit vector measurements following the von-Mises distribution and non-unit vector measurements following the isotopic Gaussian distribution, and thus extends the application scope of Theorem 3.2 in \cite{leeBayesianAttitudeEstimation2018} to include various vector measurements.

\begin{thm}\label{th:iso_postri}
	Suppose that the prior attitude distribution is a matrix Fisher distribution with parameter $F \in \mathbb{R}^{3 \times 3}$. Consider a set of mutually independent vector measurements $\left\{z_1,\cdots,z_i,\cdots,z_N,\cdots,z_{M+N}\right\} \in \left(\mathbb{R}^3\right)^{N} \times \left(\mathbb{S}^2\right)^{M}$ in $\mathcal{F_B}$ with corresponding reference vectors $\left\{r_1,\cdots,r_i,\cdots,r_N,\cdots,r_{M+N}\right\} \in \left(\mathbb{R}^3\right)^{N} \times \left(\mathbb{S}^2\right)^{M}$ in $\mathcal{F_I}$, of which the first $N$ non-unit vector measurements follows (\ref{eq:mea_model}) for some $\{Q_i=\sigma_i^{-2}I_{3 \times 3},\ 
\mathbb{R} \ni \sigma_i > 0\}^{N}_{i=1} \in \left( \mathbb{R}^{3\times 3} \right)^N$ and the remaining $M$ unit vector measurements are characterized by
\begin{gather}\label{eq:uni_mea_model}
    p(z_i) = \frac{\kappa_i}{4 \pi \sinh \kappa_i} \exp(\kappa_i r_i^T R z_i),
\end{gather}
where $\kappa_i \in \mathbb{R}$ is the concentration parameter and $\text{sinh}(\cdot)$ is the hyperbolic sine function. Then, the posterior attitude distribution for $R|\left(z_1,\cdots,z_{M+N}\right)$ is also a matrix Fisher distribution satisfying
	\begin{align}\label{eq:iso_post}
	   &R|\left(z_1,z_2,\cdots,z_N\right) \nonumber\\
       &\sim \mathcal{M}\left(F+\sum_{i=1}^{N}\sigma_i^{-2}r_iz_i^T + \sum_{i=N+1}^{M+N}\kappa_i r_iz_i^T\right),
	\end{align}
\end{thm}

\begin{pf}
	According to (\ref{eq:mea_model}) and (\ref{eq:mea_cov}), the first $N$ vector measurements obey the Gaussian distribution as
	\begin{align}
		&p\left( {z_i |\;R} \right) = \nonumber\\
		&\frac{1}
		{{\sqrt {\left( {2\pi } \right)^3 \det \left( {Q_i } \right)} }}\text{exp}\left[ { - \frac{1}
			{2}\left( {z_i  - R^T r_i } \right)^T Q_i^{ - 1} \left( {z_i  - R^T r_i } \right)} \right],
	\end{align}
 and the remaining $M$ unit vector measurements follow the von-Mises Fisher distribution as
 \begin{align}
     p(z_i|R) = \frac{\kappa_i}{4\pi \sinh{\kappa_i}} \exp\left( 
\kappa_i r_i^T R z_i \right).
 \end{align}
 
	Let $Z=\left(z_1,\cdots,z_i,\cdots,z_N,\cdots,z_{M+N}\right)$. According to Bayes' rule, we have
	\begin{equation}\label{eq:bayes_4_1}
		\pi {\text{(}}R|Z{\text{)}} = \frac{{\pi \left( R \right)p\left( {Z|R} \right)}}
		{{\int_{SO\left( 3 \right)} {\pi \left( R \right)p\left( {Z|R} \right)} }} \propto \pi \left( R \right)p\left( {Z|R} \right),
	\end{equation}
where $\pi(R)$ is the prior attitude distribution, which is assumed to be a matrix Fisher distribution. Substituting (\ref{eq:MF_de}) into (\ref{eq:bayes_4_1}), we have
	
	\begin{align}\label{eq:th4_1}
		&\pi \left( {R|\;Z} \right) \propto \underbrace {\vphantom{{\exp \left\{ { - \frac{1}
{2}\sum\limits_i {\prod\limits_{i = 1}^N {p(z_i |R)} } } \right\}}}{\text{etr}}\left( {F^T R} \right)}_A \underbrace {\prod\limits_{i = 1}^N {p(z_i |R)} }_B \underbrace {\prod\limits_{i = N+1}^{M+N} {p(z_i |R)} }_C
	\end{align}
	Conducting algebraic transformation to $B$ yields
	\begin{align}\label{eq:th4_2}
		B&\propto {\text{exp}}\left\{ { - \frac{1}
{2}\left[ {\underbrace{ {\text{tr}} \left( {\sum\limits_i^N {Q_i^{ - 1} z_i z_i^T } } \right)}_D - \underbrace{ {\text{tr}} \left( {\sum\limits_i^N {Q_i^{ - 1} z_i (R^T r_i )^T } } \right)}_E} \right.} \right. \nonumber\\ 
		&\left. {\left. { - \underbrace{ {\text{tr}} \left( {\sum\limits_i^N {Q_i^{ - 1} R^T r_i z_i^T } } \right)}_G + \underbrace { {\text{tr}} \left( {\sum\limits_i^N {Q_i^{ - 1} R^T r_i r_i^T R} } \right)}_H} \right]} \right\}
	\end{align} 
 Applying the cyclic permutation property of the trace operator and symmetry of $Q_i^{-1}$ to $G$  in (\ref{eq:th4_2}), we have
\begin{align}
    G &={ {\text{tr}} \left( {\sum\limits_i^N { z_i ( R^T r_i) ^T Q_i^{ - 1} } } \right)} \nonumber\\
      &={ {\text{tr}} \left( {\sum\limits_i^N { (Q_i^{ - 1})^T R^T r_i z_i  ^T  } } \right)}  = E.
\end{align}
Then rearranging $C$ in (\ref{eq:th4_1}), we have
\begin{align}
    C \propto \text{exp} \left( \sum\limits_{i = N + 1}^{M + N} {\kappa_i r_i^T R z_i }  \right) = \text{etr} \left( \sum\limits_{i = N + 1}^{M + N} {\kappa _i z_i r_i^T R }  \right).
\end{align}

Because the random variable in the posterior is $R$, $D$ in (\ref{eq:th4_2}) is a constant. Substitute $Q_i=\sigma^{-2}I_{3 \times 3}$ into (\ref{eq:th4_1}), then we have
	\begin{align}
	&\pi \left( {R|Z} \right) \nonumber \\
	&\propto \text{etr}\left( {F^T R} \right)\text{etr}\left[ {\left( {\sum\limits_i^N {\sigma _i^{ - 2} R^T r_i z_i^T } } \right) - } \right. \nonumber \\
	&\left. {\frac{1}
		{2}\left( {\sum\limits_i^N {\sigma _i^{ - 2} R^T r_i r_i^T R} } \right)} \right] \text{etr} \left( \sum\limits_{i = N + 1}^{M + N} {\kappa _i z_i r_i^T R }  \right)
	\end{align}
	Besides, we have 
	\begin{align}
		& \text{etr}\left[ { - \frac{1}{2}\left( {\sum\limits_i^N {\sigma _i^{ - 2} R^T r_i r_i^T R} } \right)} \right] \nonumber \\
		&= \exp \left( { - \frac{1}{2}\sum\limits_i^N {\sigma _i^{ - 2} \left\| {R^T r_i } \right\|^2 } } \right),
	\end{align}
	and because $R \in SO(3)$, $\sigma _i^{ - 2}\left\| {R^T r_i } \right\|^2 \equiv \sigma _i^{ - 2}\left\| {r_i } \right\|^2$ is constant. Additionally, using the invariance of trace operators under transpose, (\ref{eq:th4_1}) can be written as,
	\begin{align}
		&\pi \left( {R\left| Z \right.} \right) \nonumber\\
        &\propto \text{etr}\left( {F^T R} \right)\text{etr}\left( {\sum\limits_i^N {\sigma _i^{ - 2} R^T r_i z_i^T } } \right) \text{etr} \left( \sum\limits_{i = N + 1}^{M + N} {\kappa _i z_i r_i^T R }  \right) \nonumber \\
        &= \text{etr}\left( {F^T R} \right)\text{etr}\left( {\sum\limits_i^N {\sigma _i^{ - 2} z_i r_i^T  R } } \right) \text{etr} \left( \sum\limits_{i = N + 1}^{M + N} {\kappa _i z_i r_i^T R }  \right) \nonumber \\
		&= \text{etr}\left[ {\left( {F+\sum_{i=1}^{N}\sigma_i^{-2}r_iz_i^T + \sum_{i=N+1}^{M+N}\kappa_i r_iz_i^T } \right)^T R} \right], 
	\end{align}
	which verifies (\ref{eq:iso_post}).
\end{pf}

Theorem \ref{th:iso_postri} also provides a new perspective to explain the effectiveness of the SVD method \cite{markleyAttitudeDeterminationUsing1988}. When the prior distribution is a uniform distribution on $SO(3)$ and vector measurement errors follow the isotropic distribution, we show that the result of the SVD method is the same as the maximum posterior estimation (MAP) in Proposition \ref{pro:svd}.

\begin{prop}\label{pro:svd}
	Consider a set of unit and non-unit vector measurements $\left\{z_1,z_2,\cdots,z_i,\cdots,z_{M+N}\right\} \in \left(\mathbb{R}^{3}\right)^{M+N}$ in $\mathcal{F_B}$ defined as Theorem \ref{th:iso_postri}. Supposing that the prior attitude distribution is the uniform distribution on $SO(3)$, whose probability density function is given by $\pi(R) = 1$ with $R \in SO(3)$. The MAP of attitude is the same as the attitude given by the SVD method with weights given by $w_i=  {{{(\sigma _i^{ - 2} )}}/
({{\sum\nolimits_{i = 1}^N {\sigma _i^{ - 2}  + \sum\nolimits_{i = N + 1}^M {\kappa _i } } }}})$ for $i=1,\dots,N $ and $w_i=  {{{(\kappa _i )}}/
({{\sum\nolimits_{i = 1}^N {\sigma _i^{ - 2}  + \sum\nolimits_{i = N + 1}^M {\kappa _i } } }}})$ for $i=N+1,\dots,M$.

\end{prop}

\begin{pf}
	See Appendix A.
\end{pf}

\subsection{Unscented Transformation with Matrix Fisher Distributions}\label{sec:4_2}

While the application of Bayes' rule is convenient for constructing estimators with matrix Fisher distributions, the full likelihood attitude distribution can can make the posterior distribution difficult to specify mathematically when measurement models fail to satisfy the assumptions in Section 4 or \cite{leeBayesianAttitudeEstimation2018}. In particular, when vector measurement errors follow the non-isotropic Gaussian distribution, the derivation in Theorem 4.1 does not hold and the posterior attitude distribution is rewritten as
\begin{align}
	\pi \left( {R\left| Z \right.} \right) &\propto {\rm etr}\Bigg[ {\Bigg( {F + } } \nonumber \\ 
	&\left. {\sum\limits_i^N {Q_i^{ - 1} r_i z_i^T } } \right)^T R\left. { + \left( {\sum\limits_i^N { - \frac{1}
				{2}Q_i^{ - 1} R^T r_i r_i^T R} } \right)} \right],
\end{align}
which can be rewritten as a Fisher-Bingham distribution \cite{arnold2013statistics} and is characterized by a parameter matrix in $\mathbb{R}^{3 \times 3}$ and a symmetric matrix in $\mathbb{R}^{9\times 9}$. Using the Fisher-Bingham distribution to represent the uncertainty of the attitude adds complexity to both propagation and correction steps. More specifically, the additional 45 parameters (5 times more than the parameters of the matrix Fisher distribution) of the Fisher-Bingham distribution significantly increase the computational complexity of propagation, and the maximum a posteriori estimate for the Fisher-Bingham distribution can not be calculated directly by the SVD due to the extra quadratic term with respect to $R$.

In order to mitigate the computational burden caused by the complex posterior distribution, this subsection proposes a global unscented transformation to obtain an approximated likelihood probability distribution. It maintains the form of a matrix Fisher distribution even when the vector measurement errors are characterized by the non-isotropic Gaussian distributions. The detailed steps of the unscented transformation are presented as follows.

\begin{defn}\label{de:ut}
	 (Unscented Transformation) Consider a set of vector measurements $\left\{z_1,z_2,\cdots,z_i,\cdots,z_N\right\} \in \left(\mathbb{R}^{3}\right)^N$ in $\mathcal{F_B}$ and corresponding reference vectors $\left\{r_1,r_2,\cdots,r_i,\cdots,r_N\right\} \in \left(\mathbb{R}^{3}\right)^N$ in $\mathcal{F_I}$. Firstly, for each vector measurement $z_i$, a set of $2d + 1$ sigma points are given by
	\begin{align*}
		&z_i^{\sigma j}  = z_i ,{\text{(}}j = 0{\text{)}}, \\
		&z_i^{\sigma j}  = z_i  + \Delta z_i^{\sigma j} ,(j = 1,2, \ldots 2d),\\
	\end{align*} 
	where $\Delta z_i^{\sigma j}  = \left[ {\sqrt {(Nd + \kappa )Q _i } } \right]_j$ for $j = 1,2, \ldots d$ and $\Delta z_i^{\sigma j}  =  - \left[ {\sqrt {(Nd + \kappa )Q _i } } \right]_{j-n}$ for $j = n + 1, \ldots 2d$.
 
The vector $\left[ {\sqrt {(Nd + \kappa )Q _i } } \right]_j$ is the $j$-th column vector of the matrix $\sqrt {(Nd + \kappa )Q _i }$. The parameter $d \in \mathbb{R}$ is the dimensionality of vector measurements and $\kappa \in \mathbb{R}$ is dimensionless empirical parameters. In the context of attitude estimation, $d=3$.
	
Secondly, for the $j$-th sigma point of the $i$-th vector measurement, a set of vectors is constructed to obtain $R_i^{\sigma j}$, which are given by
	\begin{align*}
		&Z_i ^{\sigma j}  = \{ z_1^0 ,z_2^0 , \ldots ,z_i^0 , \ldots z_N^0 \} ,{\text{     }}({\text{i = }}0;{\text{j = }}0) \\
		&Z_i ^{\sigma j}  = \{ z_1^0 ,z_2^0 , \ldots ,z_i^j , \ldots z_N^0 \} ,{\text{     }}({\text{i = 1,2,}} \ldots N;{\text{j = 1,2,}} \ldots 2d). \\
	\end{align*}
Then, the SVD method is applied to calculate the corresponding attitude with $Z_i ^{\sigma j}$ and the reference vector set $\left\{r_i\right\}=\left\{r_1,r_2,\dots,r_N\right\}$, yielding
	\begin{align}
		USV^T &= {\sum\limits_{z_i^j \in Z_i ^{\sigma j}} {r_i (z_i^j)^T } }\label{eq:un_1}\\
		R^{\sigma j}_i &= UV^T\label{eq:un_2}
	\end{align}
	
	Finally, the first moment of attitude $R$ is approximated by
	\begin{equation}
	\widetilde E(R) = \sum\limits_i^{N} {\sum\limits_j^{2d+1} {w_i^{\sigma j} R_i^{\sigma j} } },
	\end{equation}
	where the corresponding weight $w_i^{\sigma j}$ is given by $w_i^{\sigma j}  = {\kappa }/({{\kappa  + Nd}})$ for $i=0$ and $ j=0$, and $w_i^{\sigma j}  = {1}/2({{\kappa  + Nd}})$ for $i=1,2 \ldots N,\ j= 1,2, \ldots 2d$ .
\end{defn}
	In the general unscented Kalman filter, the full distribution of the state after nonlinear transformation is approximated by a Gaussian distribution. In the proposed unscented transformation, the matrix Fisher distribution is used to approximate the full distribution such that the uncertainty of the state $R \in SO(3)$ can be represented globally and no local coordinates are introduced.

Next, we demonstrate that the first moments generated by the proposed unscented transformation are accurate up to at least the third order.
\begin{prop}\label{pro:acc}
    Consider the first moment given by the unscented transformation in  \ref{de:ut}. The accuracy of the approximated first moment $\widetilde E(R)$ is up to the 3rd order when vector measurement errors follow the Gaussian distribution.
\end{prop}
\begin{pf}
		See Appendix B.
\end{pf}
\subsection{Attitude Estimator}
We integrate Theorem \ref{th:iso_postri} and the proposed unscented transformation to obtain an attitude estimator with the matrix Fisher distribution that accommodates both unit and non-unit vector measurements. Equation (\ref{eq:MF_proga}) is used to propagate uncertainty along the attitude kinematics. The estimator is summarized in Algorithm \ref{alg1}.
\begin{algorithm}
	\caption{Bayesian Attitude Estimator}
	\begin{algorithmic}[1] 
		\Procedure{Estimation $R_k$}{} 
			\State Let $k=0$,$R \sim \mathcal{M}(F_0)$
			
			\Loop
    			\State $F_{k+1}^- =$ \Call{Propagation}{$F_{k}^+,\ \Omega_k,\ H_k$}
    			
    			\If {$\{z_{{k+1}}^i\}$ is available}
                        \State $F_{k+1}^+ =$ \Call{Correction}{$F_{k+1}^-,\ \{z_{{k+1}}^i\}$}
                        \State Obtain $U_{k+1}^+$ and $V_{k+1}^+$ with $F_{k+1}^+$ from (\ref{eq:SVD})
                        \State $R_{k+1} = U_{k+1}^+ (V_{k+1}^+)^T$
                \Else
                        \State Obtain $U_{k+1}^-$ and $V_{k+1}^-$ with $F_{k+1}^-$ from (\ref{eq:SVD})
                         \State $R_{k+1} = U_{k+1}^- (V_{k+1}^-)^T$
                \EndIf
    			\State $k=k+1$
         \EndLoop
		\EndProcedure
		
		\Statex
		
		\Procedure{$F^+=$ Correction}{$F^-,\ \{z_{i}\}$}
		\State Compute $\widetilde E(R)$ from Definition \ref{de:ut} with the subset of $\{z_{i}\}$ consisting of the vector measurements following the non-isotropic Gaussian distributions, denoted by $\{z_{i}\}_{non}$ 
		\State Apply the proper SVD of $\widetilde E(R)$ to obtain $U,\ D,\ V$ from (\ref{eq:ER})
		\State Solve (\ref{eq:ER}) for $S$ with $D$ and $F^-_{non} = F^- + USV^T$
        \State Compute $F^+$ from (\ref{eq:iso_post}) with $\{z_{i}\}/\{z_{i}\}_{non}$ and $F^-_{non}$
		\EndProcedure

        \Statex
  
	\end{algorithmic}\label{alg1}
\end{algorithm}

\section{Numerical Examples}

We compare the proposed estimator, the attitude estimator presented in \cite{leeBayesianAttitudeEstimation2018} and the MEKF method in three different cases. In all three cases, we consider the rotational motion of a 3D pendulum \cite{leeLieGroupVariational2007} as the true attitude and angular velocity. The initial attitude and angular velocity are given as $R_{true}(0) = I_{3\times 3}$ and $\Omega_{true}(0) = 4.14 \times [1,1,1] rad/s$ respectively. The attitude is assumed to be estimated with vector and angular velocity measurements at rates of 10 Hz and 50 Hz respectively. The vector measurement errors follow the Gaussian distribution with zero mean and covariance that varies by case. The white noise of angular velocity measurements is Gaussian with $H = \sigma I_{3 \times 3}$, where $\sigma = 1\  deg/\sqrt{s}$. For each case, the simulation time is $T = 60s$ and the time step is $h = 0.02s$. 
The SVD method is utilized to estimate the attitude directly from vector measurements and calculate the measurement errors. All sets of simulations run 50 times.

To compare the two proposed estimators with the Bayesian estimator in \cite{leeBayesianAttitudeEstimation2018} when using non-unit vector measurements, we use a direct method to approximate the directional distribution of measurement vector disturbed by isotropic Gaussian noise with the von-Mises Fisher distribution presented in \cite{love2003gaussian}. For a measurement vector characterized by $x \sim \mathcal{N}(\mu, \sigma I_3)$, the possibility distribution density of normalized vector measurement $x_1 = x/\Vert x \Vert$ is given by
\begin{equation}\label{eq:norm_vmf}
    p(x_1) = \frac{\kappa}{4 \pi \text{sinh} \kappa} \text{exp} (\kappa \mu_1^T x_1),
\end{equation}
where $\kappa = \Vert \mu \Vert^2 / \sigma^2$ and $\mu_1 =\mu / \Vert \mu \Vert$. For the more general situation $x \sim \mathcal{N}(\mu, \Sigma)$, it is impossible for $\kappa \in \mathbb{R}$ in the von-Mises distribution to approximate all characteristics of non-isotropic dispersion described by $\Sigma \in \mathbb{R}^{3 \times 3}$. Therefore, $\kappa$ is set as $\kappa = 3\Vert \mu \Vert^2 / \text{tr}(\Sigma)$ to approximate the average dispersion characteristics of normalized vector measurement $x_1$.

\subsection{Case $\uppercase\expandafter{\romannumeral1}$: Isotropic Measurement Noises}

\begin{figure}[htbp]
	\begin{center}
		\includegraphics[width=\linewidth]{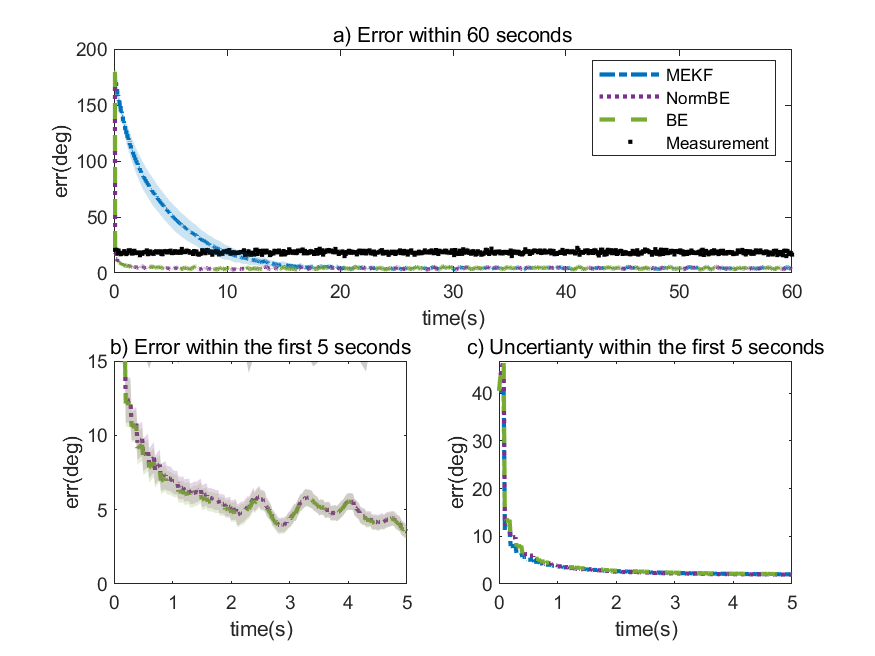}
		\caption{Case $\uppercase\expandafter{\romannumeral1}$: Isotropic Measurement Noises}\label{fig:1}
	\end{center}
\end{figure}

In this subsection, the measurement errors are chosen to be isotropic with $Q_k^i = \text{diag}[0.08, 0.08,0.08]$ and the initial parameter of the matrix Fisher distribution is
\begin{equation}
	F_0 = \text{exp}\left( \pi \hat{e}_1 \right) ,
\end{equation}
 implying the initial attitude information input to the estimators has a large error and low confidence. For the MEKF method, the corresponding degree of uncertainty is given $P = \text{diag}[\frac{1}{s_2+s_3}, \frac{1}{s_1+s_3}, \frac{1}{s_1+s_2}]=0.5I_{3 \times 3}$. The results of the Bayesian attitude estimator (BE) proposed in Section \ref{sec:4}, the Bayesian estimator presented in \cite{lee2018bayesian} with normalization (NormBE) , and the MEKF method are provided in Fig. \ref{fig:1} and Table \ref{tab:1} for comparison and the shadow in the figure represents an envelope of 95\% confidence.

\begin{table}
	\begin{center}
		\caption{Averaged error after $t=0$}\label{tab:1}
		\begin{tabular}{ccccc}
			\toprule
			& Mea.& MEKF & BE & NormBE\\
			\midrule
			est. err.($^{\circ}$) & 18.53& 15.18& 4.70& 4.71
\\
			\midrule
                time(s)& -& 0.83& 11.11&10.38\\
                \bottomrule
		\end{tabular}
	\end{center}
\end{table}

For the two estimators based on the matrix Fisher distribution, the attitude estimation errors rapidly decrease to a low level from the large initial error within 0.2 seconds, while the estimation error of MEKF descends to the same level until $t=20$ seconds. Due to the fast convergence rate, both estimators yield smaller mean attitude estimation errors than MEKF. Because NormBE needs approximate measurement uncertainty with the von-Mises Fisher distribution, which introduces more errors. Therefore, the accuracy of NormBE is slightly lower than BE. Additionally, the uncertainty of BE and NormBE decreases in a manner compatible with estimation errors, while the uncertainty of MEKF descends rapidly when the estimation error is still large. This implies that the uncertainty prediction given by MEKF is misleading.


\subsection{Case $\uppercase\expandafter{\romannumeral2}$: Non-isotropic Measurement Noises}

\begin{figure}[htbp]
	\begin{center}
		\includegraphics[width=\linewidth]{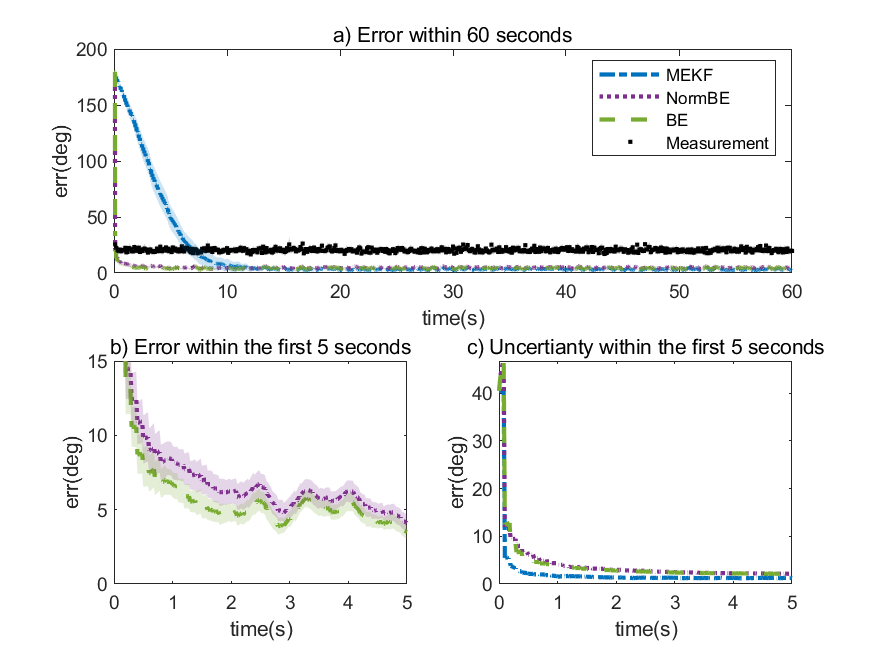}
		\caption{Case $\uppercase\expandafter{\romannumeral2}$: Non-isotropic Measurement Noises}\label{fig:2}
	\end{center}
\end{figure}

In this subsection, the vector measurement error is chosen to be non-isotropic with $Q_k^i = \text{diag}[0.01, 0.01,0.30]$ and the initial parameters are set the same as those in $Case \  \uppercase\expandafter{\romannumeral1}$. The performance of BE, NormBE, and MEKF is summarized in Fig. \ref{fig:2} and Table \ref{tab:2} for comparison.

\begin{table}
	\begin{center}
		\caption{Averaged error after $t=0$}\label{tab:2}
		\begin{tabular}{ccccc}
			\toprule
			& Mea.& MEKF & BE & NormBE\\
			\midrule
			est. err.($^{\circ}$) & 20.20& 14.08& 4.70& 5.18
\\
			\midrule
                time(s)& -& 0.85& 11.12&11.06\\
                \bottomrule
		\end{tabular}
	\end{center}
\end{table}

It can be observed that, even in the presence of non-isotropic measurement noise, the estimators with the matrix Fisher distribution outperform MEKF in terms of accuracy, similar to Case $\uppercase\expandafter{\romannumeral1}$. Meanwhile, the accuracy descending caused by normalization is more significant, due to the difficulty for the von-Mises Fisher distribution to approximate non-isotropic noise. Although the underlying posterior attitude distribution is not a matrix Fisher distribution in Case $\uppercase\expandafter{\romannumeral2}$, the proposed estimator performs well and faster convergences to smaller errors than NormBE. Moreover, the proposed estimator consumes almost the same calculation time as the previous Bayesian estimator with the matrix Fisher distribution, which implies that the estimator presented in this note is more effective in the case of non-isotropic measurement noise.

\subsection{Case $\uppercase\expandafter{\romannumeral3}$: Large Uncertainty of Angular Velocity Measurements}

\begin{figure}[htbp]
	\begin{center}
		\includegraphics[width=\linewidth]{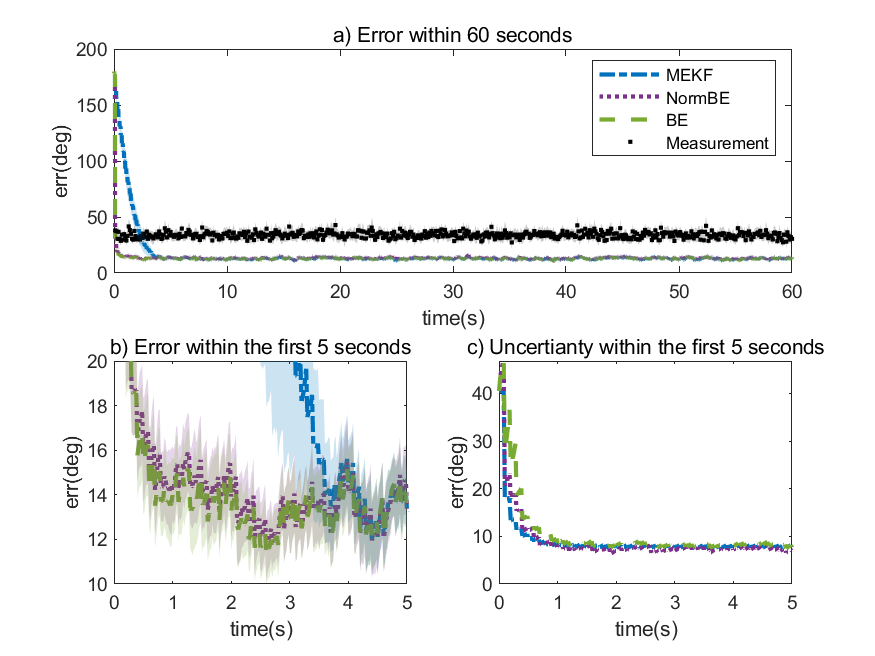}
		\caption{Case $\uppercase\expandafter{\romannumeral3}$: Large Measurement Uncertainty}\label{fig:3}
	\end{center}
\end{figure}

\begin{figure}[htbp]
	\begin{center}
		\includegraphics[width=\linewidth]{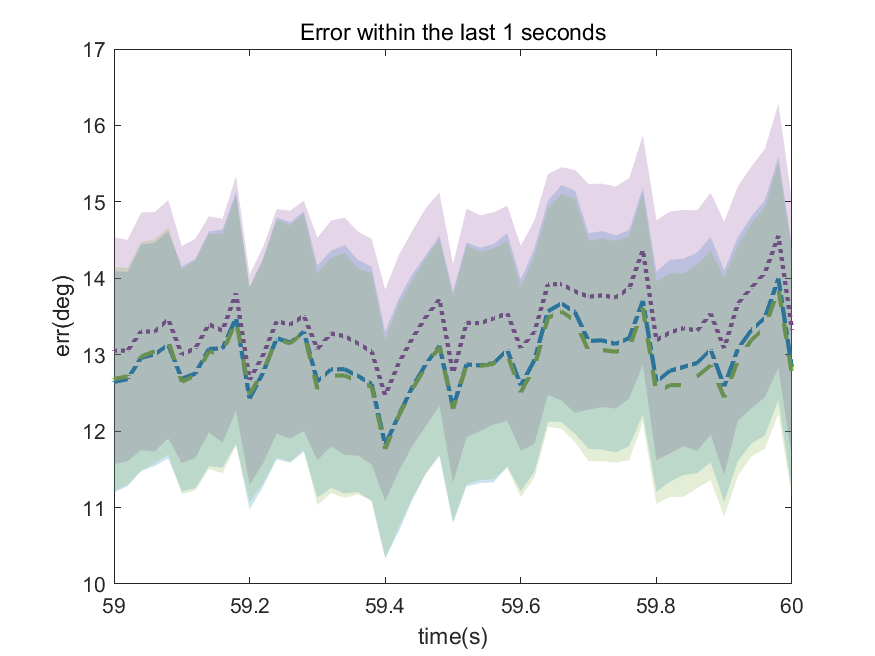}
		\caption{Case $\uppercase\expandafter{\romannumeral3}$: Steady-State Errors}\label{fig:4}
	\end{center}
\end{figure}

In this subsection, the white noise of angular velocity is Gaussian with $H = \sigma I_{3 \times 3}$, where $\sigma = 10\  deg/\sqrt{s}$ and the measurement errors are chosen to be isotropic with $Q_k^i = \text{diag}[0.24, 0.24,0.24]$. Other parameters are set the same as those in $Case \  \uppercase\expandafter{\romannumeral1}$. The angular velocity measurements are distributed with greater noise than the typical value and vector measurement uncertainty is also large, which simulates the conditions that measurements are subject to strong interference. The performance of BE, NormBE, and MEKF are summarized in Fig. \ref{fig:3}, Fig. \ref{fig:4} and Table \ref{tab:3} for comparison and the shadow in the figure represents a 95\% confidence interval.

\begin{table}
	\begin{center}
		\caption{Averaged error after $t=0$}\label{tab:3}
		\begin{tabular}{ccccc}
			\toprule
			& Mea.&MEKF &BE& NormBE \\
			\midrule
			est. err.($^{\circ}$)& 33.92&16.19& 13.19& 13.59
\\
			\midrule
                time(s)& -&0.82&8.00& 7.96\\
                \bottomrule
		\end{tabular}
	\end{center}
\end{table}

In $Case \  \uppercase\expandafter{\romannumeral3}$, the convergence of MEKF accelerates compared with $Case \  \uppercase\expandafter{\romannumeral1}$ but is still slower than the proposed estimator and NormBE. The reason is that the confidence in wrong attitude estimation decreases more rapidly during propagation than $Case \  \uppercase\expandafter{\romannumeral1}$ due to larger angular velocity measurement noise. Therefore, the wrong information propagated from the initial state with large error is filtered out much faster. Although MEKF acquires a significant promotion in convergence rate due to the large uncertainty of angular velocity measurements, the proposed estimator maintains advantages over MEKF in both convergence rate and accuracy. Meanwhile, the proposed estimator is still more accurate than the estimator with normalization. The Fig. \ref{fig:4} demonstrates that these advantages are also related to smaller steady-state errors of the proposed estimator compared with NormBE.

\section{Conclusion}
In this note, we develop a Bayesian attitude estimator with matrix Fisher distribution that can accommodate both unit and non-unit vector measurements without normalization. A new theorem and a novel unscented transformation are presented to deal with non-unit vector measurements with isotropic and non-isotropic Gaussian noise, respectively. The proposed estimator exhibits advantages compared with the previous Bayesian estimator \cite{leeBayesianAttitudeEstimation2018} with normalization and the classic MEKF in three different simulation examples. Moreover, the proposed theory and novel unscented transformation allow to propagate the uncertainty from $\mathbb{R}^3$ to $SO(3)$, and are fundamental contributions that can be applied to more generic estimation problems.

%

\bibliographystyle{unsrt}        
\bibliography{autosam}        

%
\appendix
\section{Appendix A: Proof of Proposition \ref{pro:svd}}

When the prior attitude distribution is a uniform distribution on $SO(3)$, the probability density function of posterior attitude distribution is
\begin{equation}
	p\left( R \right) = \frac{1}
	{{c\left( F \right)}}etr\left[ {F^T R} \right],
\end{equation}
where parameter $F$ is given by
\begin{equation}
	F = \sum\limits_i^N {\sigma _i^{ - 2} r_i z_i^T }.
\end{equation}
According to properties of the matrix Fisher distribution on $SO(3)$, the MAP is yielded by
\begin{equation}
	R_{MAP} = U_FV_F^T
\end{equation}
where $U_F$ and $V_F$ are obtained by proper SVD of $F$.

On the other hand, for SVD method, the Wahba loss function in \cite{wahba1965least} is
\begin{equation}
	L(R) = 1-tr(B^TR)
\end{equation}
where $B=\sum_{i=1}^{N}{w _i r_i z_i^T }$.
The proper SVD is given by
\begin{equation}
	B=kF=kU_FS_FV_F^T=U_FS_BV_F^T,
\end{equation}
where $k={\sum\nolimits_{i = 1}^N {\sigma _i^{ - 2}  + \sum\nolimits_{i = N + 1}^M {\kappa _i } } }$. Thus, the attitude estimation given by SVD method is
\begin{equation}
	R_{SVD}=U_FV_F^T,
\end{equation}
which is same as the result given by MAP.

\section{Appendix B: Proof of Proposition \ref{pro:acc}}

Noting that $Z_0  = [z^T _1 ,z_2^T , \ldots z_m ^T ]^T $ denotes vector measurements without noise, the corresponding attitude matrix is given by (\ref{eq:un_1}) and (\ref{eq:un_2}) in section 5 and it is abbreviated as
\begin{equation}
	R = f(Z_0)
\end{equation}
Noting that the actual measurement result is $Z=Z_0+\delta Z$ and applying differential operator $D_{\delta z} f$ yields
\begin{equation}
	D^i _{\delta Z} f = (\delta Z \cdot \nabla )^i f(Z_0 )
\end{equation}
where $\left(\cdot \right)$ is vector dot product. The Taylor series expansion of $f(Z)$ at $Z=Z_0$ is given by
\begin{align}
	f(Z)&= f(Z_0 ) + D_{\delta Z} f + \frac{1}
	{2}D^2 _{\delta Z} f + \frac{1}
	{3}D^3 _{\delta Z} f +  \cdots \nonumber \\
	&= f(Z_0 ) + \left. {\frac{{\partial f}}
		{{\partial Z^T }}} \right|_{Z = Z_0 } \delta Z + \nonumber \\
	&\frac{1}
	{2}(\nabla ^T \delta Z\delta Z^T \nabla ^T )f(Z_0 ) +  \cdots 
\end{align}
Vector measurement errors are assumed to follow Gaussian distribution, which is centrally symmetric. Therefore, the first moment of $f(Z)$ is computed as
\begin{align}
	E\left[ {f(Z)} \right] &= f(Z_0 ) + \left. {\frac{{\partial f}}
		{{\partial Z^T }}} \right|_{Z = Z_0 } E(\delta Z) + \nonumber \\ 
	&\frac{1}
	{2}(\nabla ^T E(\delta Z\delta Z^T )\nabla ^T )f(Z_0 ) + E\left( {\frac{1}
		{3}D_{\delta Z}^3 f +  \cdots } \right) \nonumber \\ 
	&= f(Z_0 ) + \frac{1}
	{2}(\nabla ^T E(\delta Z\delta Z^T )\nabla ^T )f(Z_0 ) + \nonumber \\ 
	&E\left( {\frac{1}
		{{4!}}D_{\delta Z}^4 f + \frac{1}
		{{6!}}D_{\delta Z}^6 f \cdots } \right) \nonumber \\ 
	&= f(Z_0 ) + \frac{1}
	{2}(\nabla ^T P_{ZZ} \nabla ^T )f(Z_0 ) + \nonumber \\ 
	&E\left( {\frac{1}
		{{4!}}D_{\delta Z}^4 f + \frac{1}
		{{6!}}D_{\delta Z}^6 f \cdots } \right)
\end{align}
where $P_{zz}$ is covariance matrix of $Z$. Without loss of generality, vector measurements are assumed to be independent of each other, and thus $P_{zz}$ is a block diagonal matrix as follows:
\begin{equation}
	P_{zz}  = diag\left[ {\Sigma _1 ,\Sigma _2 , \cdots ,\Sigma _m } \right] = \left[ {\begin{array}{*{20}c}
			{\Sigma _1 } & {0_{3 \times 3} } &  \cdots  & {0_{3 \times 3} }  \\
			{0_{3 \times 3} } & {\Sigma _2 } & {} & {0_{3 \times 3} }  \\
			\vdots  & {} &  \ddots  &  \vdots   \\
			{0_{3 \times 3} } & {0_{3 \times 3} } &  \cdots  & {\Sigma _m }  \\
			
	\end{array} } \right].
\end{equation}

Noting that $\delta Z_i^{\sigma j}  = [0_{3 \times 1} ^T ,0_{3 \times 1} ^T , \cdots (\Delta z_i^{\sigma j} )^T , \cdots 0_{3 \times 1} ^T ]$, the matrix $R_i^{\sigma j} $ is given by 
\begin{equation}
	R_i^{\sigma j}  = f(Z_0  + \delta Z_i^{\sigma j} )
\end{equation}
The Taylor series expansion of $R_i^{\sigma j}$ at $Z_0$ is 
\begin{align}
	&f(Z_0  + \delta Z_i^{\sigma j} ) = f(Z_0 ) + \left. {\frac{{\partial f}}
		{{\partial Z^T }}} \right|_{Z = Z_0 } \delta Z_i^{\sigma j}  +  \hfill \nonumber\\
	&\frac{1}
	{2}\left[ {\nabla ^T \delta Z_i^{\sigma j} (\delta Z_i^{\sigma j} )^T \nabla ^T } \right]f(Z_0 ) +  \cdots  \hfill \nonumber\\
	&= f(Z_0 ) + \left. {\frac{{\partial f}}
		{{\partial Z^T }}} \right|_{Z = Z_0 } \delta Z_i^{\sigma j}  \hfill \nonumber\\
	&+ \frac{1}
	{2}\left\{ {\nabla ^T diag\Bigg[0_{3 \times 3} , \cdots \left[ {\sqrt {(mn + \kappa )\Sigma _i } } \right]_j \left[ {\sqrt {(mn + \kappa )\Sigma _i } } \right]_j^T ,} \right. \hfill \nonumber\\
	&\cdots \Bigg]\nabla ^T \Bigg\} f(Z_0 ) +  \cdots  \hfill \nonumber\\ 
\end{align}
Because the sigma point is selected symmetrically and
\begin{equation}
	\sum\limits_j {\left[ {\sqrt {(mn + \kappa )\Sigma _i } } \right]_j \left[ {\sqrt {(mn + \kappa )\Sigma _i } } \right]_j ^T }  = 2(mn + \kappa )\Sigma _i,
\end{equation}
the $\widetilde E(R)$ in Definition 3 is 
\begin{align}
	&\widetilde E(R) \nonumber\\
	&= \sum\limits_i {\sum\limits_j {w_i^{\sigma j} R_i^{\sigma j} } }  \hfill \nonumber\\
	&= \frac{\kappa }
	{{\kappa  + mn}}f(Z_0 ) + \sum\limits_i {\frac{{2n}}
		{{2(mn + \kappa )}}f(Z_0 )}  +  \hfill \nonumber\\
	&\sum\limits_i {\frac{1}
		{2}\left\{ {\nabla ^T diag[0_{3 \times 3} ,0_{3 \times 3} , \cdots ,\Sigma _i , \cdots ,0_{3 \times 3} ]\nabla ^T } \right\}f(Z_0 ) +  \cdots }  \hfill \nonumber\\
	&= f(Z_0 ) + \frac{1}
	{2}\left\{ {\nabla ^T diag[\Sigma _1 ,\Sigma _2 , \cdots ,\Sigma _m ]\nabla ^T } \right\}f(Z_0 ) +  \cdots  \hfill \nonumber\\
	&= f(Z_0 ) + \frac{1}
	{2}\left\{ {\nabla ^T P_{ZZ} \nabla ^T } \right\}f(Z_0 ) +  \cdots  \hfill,
\end{align}
whose first four terms are exactly the same as (B.4).

\end{document}